\documentstyle[preprint,aps]{revtex}
\begin{document}
\draft
\date{ 15 March 1993 }
\title{Surfactant-Mediated Growth of Nonequilibrium Interfaces}
\author{Albert-L\'aszl\'o Barab\'asi$^*$}
\address{Center for Polymer Studies and Department of Physics,
 Boston University, Boston, MA 02215 USA }
\maketitle
\begin{abstract}

        A number of recent experiments have showed that surfactants
can  modify the growth mode of an  epitaxial film, suppressing islanding and
promoting layer-by-layer growth. Here   I introduce
 a set of  coupled  equations to describe  the nonequilibrium roughening
of an interface covered with a thin surfactant layer. The
surfactant may drive the system into  a
 novel phase, in which the surface roughness is negative,
corresponding to a flat surface.

\end{abstract}

\pacs{PACS numbers: 68.55 -a, 68.35 Fx, 64.60 Ht}
\narrowtext
There is  much theoretical interest in the statistical properties of
nonequilibrium surfaces.
The growing interfaces naturally evolve into self-affine structures;
the surface morphology and the dynamics of roughening exhibit
simple scaling  behavior despite the complicated nature of the growth process
\cite{VIC}.
In particular, much attention has  focused on
different models  to describe thin-film growth by molecular-beam epitaxy (MBE)
\cite{WV,VILLAIN,GB,LS,TN}.

Under ideal MBE conditions the primary relaxation mechanism is  surface
diffusion,  which  conserves the mass of the film.
Experimentally   both  lattice strain and surface free energy
determine whether the film undergoes layer-by-layer growth, islanding,
or layer-by-layer growth followed by islanding. In experiments involving
 growth of Ge on Si(100) surface  layer-by-layer growth is limited to 3-4
monolayers (ML)
due to the lattice mismatch between
 Si and Ge
and is followed by formation of unstrained Ge islands.  It
 was  shown recently that  islanding in the Ge/Si system can be
suppressed effectively by use of a surfactant monolayer, {\it changing the
growth mode} from island growth to layer-by-layer growth  \cite{COPEL}.
Suitable surfactants  such as As and Sb  strongly reduce
 the surface free energy of both Si and Ge surfaces and
 segregate at the surface during growth.
A number of subsequent experiments showed that surfactants can change
the surface morphology in a wide variety of systems: Sb altered the
 growth behavior of Ag on Ag(111) \cite{Ag}, antimony was found to change the
structure of islands in Ge/Si growth \cite{antimony} and  Te was used as
surfactant to
 sustain layer-by-layer growth of  InAs   on GaAs(001) \cite{Te}.

Microscopically, arsenic placed on the surface of Si or Ge lowers the surface
tension  by eliminating the dangling bond states. But for ordinary surfactants
reduced surface tension {\it enhances} roughness. The reduced adatom diffusion
might change the growth mode, but so far most  continuum models with or without
surface diffusion predict rough  interfaces \cite{WV,VILLAIN,GB,LS,TN,SP}.
While  the microscopic mechanism
of  As and Ge/Si interaction is much investigated \cite{TIM},
 understanding in
the framework of continuum growth models is still lacking.

In this paper I  study
the generic problem of nonequilibrium roughening
of an interface covered by a thin surfactant layer (see Fig \ref{fig1}).
 Building on
 experimental results
and general symmetry principles,  a set of nonequilibrium
equations are proposed to describe the growth of an interface coupled to the
fluctuations
in the surfactant coverage. The analytic study of these equations
indicates  that
the surfactant changes drastically the morphology of the interface in
2+1 dimensions. In  particular,  the coupled
system supports the existence of  a novel phase
characterized  by negative roughness exponent, which  can be  identified with
a morphologically flat surface.

As mentioned above, under ideal MBE conditions, relaxation proceeds via
surface diffusion.  In contrast to the ideal MBE,
there is experimental evidence that surfactant mediated growth of Ge on Si
proceeds by highly local Ge incorporation with minimum surface diffusion
\cite {TR}.   Ge atoms that adhere to the As-capped surface  rapidly
exchange sites with the As atoms and incorporate into  subsurface sites.
In the absence of surface diffusion, the growth equation may  contain terms
which violate mass conservation \cite{CONS}. The simplest nonlinear growth
equation
with nonconserved dynamics was introduced by
 Kardar, Parisi, and Zhang (KPZ) \cite{KPZ}:
\begin{equation}
\partial_t h = \nu \nabla^2h + \lambda ({\bf \nabla} h)^2 + \eta.
\label{KPZ}
\end{equation}
Here  $h(x,t)$ is the height of the interface in $d=d'+1$ dimensions.
The first term on the right hand side describes relaxation of the surface by
a surface  tension $\nu$. The second term is the lowest order nonlinear term
that can appear in the interface growth equation, and  is related to
lateral growth.
$\eta(x,t)$ is a stochastic noise driving the growth; it can  describe
thermal and beam intensity fluctuations.
Additional terms in (\ref{KPZ}) will  include  the coupling
 to the surfactant fluctuations.

	An efficient surfactant must fulfill two criteria: it must be
sufficiently mobile to avoid incorporation, and it must surface segregate.
Careful experimental studies showed for the Ge/Si system that the bulk
As concentration is less than 1\%; thus the effect of As on growth is
 a surface phenomena \cite{COPEL}. Neglecting the desorption of the surfactant
atoms,
 the equation governing the
surfactant kinetics  obeys mass conservation.
This leads   to the continuity  equation $\partial_t v = - {\bf \nabla}
\cdot    {\bf j}
 + \eta'$, where $v(x,t)$ is the {\it width of the surfactant layer}
\cite{WIDTH}, $\eta'$ is
a  conserved uncorrelated noise which incorporates the random local
fluctuations of the surfactant, and  ${\bf j}$ is the particle-number
current density.
 The simplest linear
equation with conserved dynamics correctly incorporating the effect
of surface
diffusion is \cite{WV}
\begin{equation}
\partial_t v = - K \nabla^4 v + \eta'. \label{DIFF}
\end{equation}

To account for the coupling between the growing surface and the surfactant
it is necessary  to introduce additional terms in Eq.  (\ref{KPZ})  and
 (\ref{DIFF}). There
are two main criterias which restrict our choice: The coupling terms must
satisfy the symmetry conditions characteristic of the interface and the
obtained set of equations should be self-consistent, i.e. the resulting
dynamics should not generate further  nonlinear terms. In addition
 the coupling terms included in Eq.  (\ref{DIFF}) must  obey the
required mass conservation for the surfactant.

	The simplest set of equations that  satisfy the above conditions is

\begin{mathletters}
\label{COUPL}
\begin{equation}
\partial_t h = \nu \nabla^2h + \lambda ({\bf \nabla} h)^2 + \beta
({\bf \nabla} v)^2 +\eta_0   \label{COUPL.a}
\end{equation}
\begin{equation}
\partial_t v = - K \nabla^4 v + \gamma \nabla^2 [({\bf \nabla} h) \cdot
( {\bf \nabla} v)] + \eta_1,
\label{COUPL.b}
\end{equation}
\end{mathletters}
where the noise terms  $\eta_0$ and $\eta_1$ are assumed to be Gaussian
distributed
with zero mean and the following correlator:
\begin{equation}
<\eta_i(x,t)\eta_i(x',t')>= {\cal D}_i \delta(x-x')\delta(t-t').
\end{equation}
Here
\begin{equation}
{\cal D}_0=D_0
\end{equation}
and
\begin{equation}
{\cal D}_1=-D_1 \nabla^2 + D_2 \nabla^4. \label{D2}
\end{equation}

The $D_2$ term  is generated by $D_0$ and $D_1$ as will be shown below.

The generic nonlinear term $({\bf \nabla} v)^2$ in (\ref{COUPL.a})
can be derived using symmetry principles. In (\ref{COUPL.b}) the
$\nabla^2 [( {\bf \nabla} h) \cdot ( {\bf \nabla} v)]$ term results from
a current ${\bf j} = - {\bf \nabla}
[({\bf \nabla} h) \cdot ( {\bf \nabla} v)]$, and obeys mass conservation.
Geometrical interpretation
\cite{LS} of this term suggests that a positive $\gamma$  drives the
surfactant to cover uniformly   the irregularities of the surface, i.e.
enhances the wetting properties \cite{GENNE}. A
negative $\gamma$ has the opposite effect, assigning a non-wetting character
to the surfactant. Since in experiments there is no evidence of surfactant
agglomeration (non-wetting character), but it is energetically favorable
to terminate the Ge layer with As atoms, we assume that the  surfactant
  wets the surface, thus $\gamma > 0$.

The quantity of main  interest is the dynamic scaling of the
fluctuations characterized by the  width
$w_0^2(t,L)=~<[h(x,t)-{\overline h(t)}]^2>~ = L^{2 \chi_0} f(t/L^{z_0})$, where
$\chi_0$ is the roughness exponent for the interface $h(x,t)$, and the
dynamic exponent $z_0$  describes the scaling of the relaxation times
 with the system size $L$; $\overline h(t)$ is the mean height
of the interface at time $t$ and the $<>$  denotes ensemble
 and space average. The scaling function  $f$ has the properties
$f(u \to 0) \sim u^{2z_0/\chi_0}$ and $f(u \to \infty) \sim$const.
 In a similar way one can define $\chi_1$and $z_1$ to characterize
the fluctuations in the surfactant coverage $v(x,t)$.

 	For $\beta=0$, Eq. (\ref{COUPL.a}) reduces to the KPZ equation (\ref{KPZ}).
 For the
physically relevant dimension, $d=2+1$, the scaling exponents are not known
exactly. Extensive numerical simulations give $\chi_0= 0.385  \pm 0.005$ and
$z_0 \simeq 1.6  $ \cite{NUMER}. Thus the interface is rough and the roughness
increases with time as $w_0(t) \sim t^{\chi_0/z_0}$.

	For $\gamma=0$, Eq. (\ref{COUPL.b}) is the fourth order linear
diffusion
 equation with conserved noise (\ref{DIFF}),  which can be solved
exactly, resulting in   $z_1=4$
and $\chi_1=0$  \cite{SGG}.

Thus, neglecting the coupling terms, Eq. (\ref{COUPL.a})
and (\ref{COUPL.b})
predict rather different values   for $z$ and the roughness exponents
$\chi_i$. To see  how  the couplings change this behavior,
I have investigated Eq. (\ref{COUPL}) using
a   dynamical renormalization-group  (DRG) scheme.
During the DRG calculations only  one dynamic exponent $z=z_0=z_1$
was used, valid if the system does not decouple.
 The fast modes are integrated out in the momentum shell
$e^{-l}\Lambda_0 \leq |k| \leq \Lambda_0$, and the variables are rescaled as
$x\to e^l x$, $t \to e^{zl} t$, $h \to e^{\chi_0 l}h$, and $v\to e^{\chi_1l}v$.
The calculations have been performed up to one-loop order.

	The diagrams contributing to $\lambda$ cancel each  other, resulting in
 the scaling relation
\begin{equation}
z+\chi_0 = 2.
\label{GALILEI}
\end{equation}
 This relation is known to be  the  property of the KPZ equation  and it is a
consequence of  Galilean invariance (GI). Since the DRG conserves the GI,
 this scaling law is expected to remain valid to all orders of the perturbation
theory.

A second scaling relation  can be obtained from the
non-renormalization of the diffusion coefficient $D_1$:
\begin{equation}
z-2\chi_1-d'-2=0.
\label{D1}
\end{equation}
 The diagrams that
contribute to $D_1$ have a prefactor proportional to ${\bf k}^4$, thus they
are irrelevant (${\bf k}$ is the wave vector in the Fourier space).
 They in fact  contribute to $D_2$,
 justifying its introduction in (\ref{D2}).

	These two scaling relations already indicate that the coupled
interface/surfactant system  is qualitatively different from the  uncoupled
one. For a planar interface ($d'=2$)  (\ref{GALILEI},\ref{D1}) give
\begin{equation}
\chi_0 + 2\chi_1 = -2,
\end{equation}
which means that at least one of the exponents has to be negative.

	A third scaling relation unfortunately is not available, but
insight can be obtained from   numerical
integration of  the  flow equations obtained from the DRG
\cite{FLOW}.
The integration showed the existence of two main  regimes:

(i) In the first  regime  one or  both  of the coupling terms ($\beta, \gamma$)
 scale to zero. In this case the two equations become completely (both
 coupling terms vanish) or   partially (only one coupling term vanishes)
decoupled, and the two equations might support different dynamic exponents $z$.
The  DRG scheme used   is not reliable in this regime.

(ii) The presence of a strong coupling fixed point is expected when both of
the nonlinear terms diverge. The integration shows that this {\it coupled
phase } exists only for $z \ge 8/3$ \cite{INSTAB}.  It is important to note
that although
there is no identifiable fixed point, in this phase the scaling relations
(\ref{GALILEI}, \ref{D1}) are exact. According to (\ref{GALILEI}), for $z \ge
8/3$
the roughness exponent of the interface $\chi_0$ is negative.
With a negative roughness exponent,
every noise-created irregularity is smoothed out by the growth dynamics
and
the resulting surface becomes flat.
Thus the coupling
of the surfactant  to the growing interface results
 in the {\it suppression of the surface roughness}.
This corresponds  exactly to the
  experimentally  observed behavior, i.e.
the addition of the surfactant suppresses  islanding, resulting
 in a morphological transition from rough (without surfactant)
 to flat  (with surfactant) interface.

The roughness exponent of the surfactant from (\ref{D1}) is negative if
$z < 4$, while for $z > 4$ it becomes positive.
  In the Ge/Si system, for example,  the As
 has a saturation coverage of 1ML, which is   independent
 of the system size and is governed only  by the microscopic bonding of the
As to the Ge dangling bonds. One expects no relevant fluctuations
in the thickness of the coverage; this requires  a negative roughness exponent
for the surfactant and  thus limits the dynamic exponent to values
smaller than four.

	The DRG analysis fails to provide  the exact value of the
dynamic exponent $z$. As in the case of many other growth phenomena, simple
discrete models might be very helpful to obtain its value \cite{MODELS}.
Summarizing
the results of the direct integration of the DRG equations, for $z > 8/3$
the existence of a strong coupling fixed point is observed, in which
the interface roughness exponent is negative, corresponding to a flat phase.
There is no upper bound in $z$ for the existence of this phase, but physical
considerations suggest that $z < 4$, in order to allow the uniform surfactant
coverage observed experimentally.

It is important to note that introducing  Eq. (\ref{COUPL}) I did not
use directly the existence  of the strain due to the lattice mismatch.
Although  an important problem \cite{STRESS}, a continuum
description of strain-induced roughening is still missing. The proposed model
is expected to describe  the coupled surfactant/interface system, but
decoupling the surfactant does not necessary result in an equation describing
heteroepitaxial islanding. Further studies are necessary to understand the
microscopic (perhaps strain induced) origin of the nonlinear coupling terms.

	In  Eq. (\ref{COUPL})  the desorption of
 the surfactant atoms is neglected by considering that  (\ref{COUPL.b}) obeys
mass
conservation. Lifting the conservation law, (\ref{COUPL.b}) should be replaced
by a nonconservative equation.
Such a system has been recently studied \cite{DENIZ,alb},
 and it was found
that in most cases the coupling does not change the KPZ scaling exponents.
Enhancement of the exponents is possible {\it only} when the coupling is
one-way, i.e. one of the equations  decoupled from the other one is
  acting as  source of correlated noise.

Further linear and/or nonlinear terms added to (\ref{COUPL}) might
influence the dynamics of the system. The goal here was to derive the
{\it simplest} set of equations predicting the experimentally observed
morphological phase transition; the study of other possible nonlinear
terms and their relevance is left for future work.

In conclusion, I have introduced a set of equations to describe the
interaction of a growing surface with a surfactant. The main experimentally
motivated  requirements for  (\ref{COUPL}) were: (a) no surface diffusion
of the newly landed adatoms; (b) conservative and diffusive surfactant
dynamics, originating
from neglecting incorporation and desorption of the surfactant during the
growth process. The obtained
equations indicate the existence of a coupled phase, in which two scaling
relations between the three exponents are available. In this phase, the
roughness
exponent of the interface is negative, morphologically corresponding to a
flat interface, as observed experimentally.

 Moreover, Eq. (\ref{COUPL}) serve as a good starting point for future
 studies of an interface coupled to a local {\it conservative} field, a
 problem of
major interest in the context of recent efforts to understand the general
properties of  nonequilibrium stochastic systems.

I wish to thank M. Gyure and E. Kaxiras for useful discussions and comments on
the
manuscript and H.E. Stanley for continuous encouragement and support.
  This research was partially funded by the Hungary-USA
exchange
program of the Hungarian Academy of Sciences and the
 National Science
Foundation. The Center for Polymer
Studies is supported by National Science Foundation.

\begin{figure}
\caption{Schematic illustration of the studied surfactant/surface system.
The figure represents a cross section of the two
dimensional surface of heigh $h(x,t)$ covered by a  thin surfactant
layer
with thickness $v(x,t)$.  A newly arriving atom penetrates the
surfactant and  is deposited on the top of the growing interface
$h(x,t)$. }
\label{fig1}
\end{figure}

\begin{references}
\bibitem{CONS} In fact, even for systems with surface diffusion but
without mass conservation the long time  behavior is described by the
 KPZ equation \cite{KPZ}.  But if surface diffusion is absent, the
growth equation can have only nonconservative dynamics, as is the case
with the surfactant covered Ge interface.
\bibitem{WIDTH} Throughout this paper is  assumed that the surfactant
layer is very thin, thus nonlocal effects do not contribute to the
dynamics.    The typical experimental coverage,
 which  is the spatial average of $v(x,t)$, is around  1ML.
For coverages snaller than 1ML holes might appear in the surfactant
layer. Since the to be proposed growth equations do not depend in an
explicit form of the thickness of the layer, but only on its spatial
derivative,
the system remains  well defined even in the presence of such a holes.
\bibitem{FLOW} A  correct flow must
not scale the nonlinear terms $\beta$ and $\gamma$ to zero,
which would decouple the Eq. (\ref{COUPL.a}) and (\ref{COUPL.b}).
  The finiteness of the nonlinear terms guarantee the
validity of the scaling relations (\ref{GALILEI},\ref{D1}) as well.
\bibitem{INSTAB} The coupled phase is stable against small fluctuations
in the coefficients
and exists in a finite region of the parameter space. Since under
experimental conditions small fluctuations in the value of the
control parameters are always expected,  the stability of the system
against them ensures the persistence of the coupled phase. But large
deviations of the parameters introduce instabilities, which results
in the breakdown of the smooth phase. This is in accord with experimental
observation, that surfactant induced layer-by-layer  growth develops only under
well
controlled experimental conditions.
\end{references}
\end{document}